\documentclass[conference]{IEEEtran}
\IEEEoverridecommandlockouts
\usepackage{cite}
\usepackage{amsmath,amssymb,amsfonts}
\usepackage{graphicx}
\usepackage{textcomp}
\usepackage{xcolor}
\usepackage[a4paper, total={210mm,297mm}, top=19mm, bottom=43mm, left=13mm, right=13mm,]{geometry}
\def\BibTeX{{\rm B\kern-.05em{\sc i\kern-.025em b}\kern-.08em
    T\kern-.1667em\lower.7ex\hbox{E}\kern-.125emX}}
    
\usepackage{booktabs}
\usepackage{multirow}
\usepackage{algorithm}
\usepackage[noend]{algpseudocode}
\usepackage{blindtext}
\usepackage{graphicx}
\usepackage{wrapfig}
\usepackage[caption=false]{subfig}
\usepackage[font=footnotesize]{caption}

\usepackage{tabularx}
\usepackage{soul}
\usepackage{tablefootnote}

\usepackage{enumitem}

\usepackage{url}

\usepackage{listings}
\usepackage{xcolor}

\definecolor{mygreen}{rgb}{0,0.6,0}
\definecolor{mygray}{rgb}{0.5,0.5,0.5}
\definecolor{mymauve}{rgb}{0.58,0,0.82}
\definecolor{backcolour}{rgb}{0.95,0.95,0.92}

\begin{document}

\newcommand{\TecName}{SVAgent} 
\title{\TecName: AI Agent for Hardware Security Verification Assertion}



\author{
Rui Guo, Avinash Ayalasomayajula, Henian Li, Jingbo Zhou, Sujan Kumar Saha, Farimah Farahmandi\\
Department of Electrical and Computer Engineering, University of Florida\\
\{guor, ayalasomayajul.a, henian.li, jingbozhou, sujansaha\}@ufl.edu, farimah@ece.ufl.edu}

\maketitle

\begin{abstract}

Verification using SystemVerilog assertions (SVA) is one of the most popular methods for detecting circuit design vulnerabilities. However, with the globalization of integrated circuit design and the continuous upgrading of security requirements, the SVA development model has exposed major limitations. It is not only inefficient in development, but also unable to effectively deal with the increasing number of security vulnerabilities in modern complex integrated circuits. In response to these challenges, this paper proposes an innovative SVA automatic generation framework \TecName. \TecName~introduces a requirement decomposition mechanism to transform the original complex requirements into a structured, gradually solvable fine-grained problem-solving chain. Experiments have shown that \TecName~can effectively suppress the influence of hallucinations and random answers, and the key evaluation indicators such as the accuracy and consistency of the SVA are significantly better than existing frameworks. More importantly, we successfully integrated \TecName~into the most mainstream integrated circuit vulnerability assessment framework and verified its practicality and reliability in a real engineering design environment.

\end{abstract}


\section{Introduction} 
\label{sec:introduction}

To meet the increasing demand for consumer electronics, the current integrated circuits supply chain is shifted to a horizontal model where numerous untrusted entities need to be involved for manufacturing and testing purposes~\cite{guo2023evolute}. In this case, various security vulnerabilities might be introduced and need to be addressed urgently. Assertion-based verification is one of the most effective methods to detect hardware security vulnerabilities~\cite{9000170,farahmandi2018formal}. However, traditional verification methods rely on manual code reviews and simulation-based testing, which is not only labor-intensive but also struggles to keep pace with increasingly complex designs and evolving security threats. Existing research on the generation of SVA mainly focuses on the functionalities, while research on SVAs for detecting potential security vulnerabilities is extremely limited. Functional errors can usually be effectively detected by traditional simulation and testing methods, and their characteristics often manifest as obvious failures or anomalies in normal operation. In contrast, hardware security vulnerabilities are inherently hidden and complex. They usually do not appear in standard functional tests but require specific trigger conditions and sequences to be activated. These security vulnerabilities may include information leakage paths or permission bypass mechanisms, which may be completely invisible under normal operating conditions but may lead to catastrophic security consequences under certain conditions. Therefore, SVA for hardware security requires a deep understanding of attacker thinking patterns and possible attack vectors, which significantly increases the workload for verification engineers due to the need for manual analysis, extensive threat modeling, and precise property definition. In the meantime, this security-oriented formal verification approach is crucial to preventing such changes in functionality and information leakage~\cite{ernst1999dynamically}, especially in the current environment of growing hardware supply chain security threats.

Automatically generating source code based on natural language instructions has been proven to significantly improve programming efficiency~\cite{xu2022ide}. Developing SVA based on security requirements is essentially a complex language processing task that requires a deep understanding of the semantics of the requirements and converting them into formal specifications. Recent studies have shown that large language models (LLMs) have demonstrated outstanding capabilities in multiple dimensions, including natural language understanding, basic arithmetic reasoning, common sense reasoning, and symbolic logical thinking, and are gradually expanding to more advanced cognitive activities such as analogical reasoning and multimodal reasoning~\cite{imani2023mathprompter, qin2023chatgpt, webb2023emergent, driess2023palm}. Many AI agent frameworks can achieve very good results by using only pre-trained LLMs as the core engine and performing structured prompts~\cite{liu2023dynamic, wang2024executable}. In this process, the prompts input to the LLM are crucial. However, studies show that for SVA generation tasks, current LLM-based frameworks cannot generate SVA with high accuracy~\cite{kande2023llm,divas2023}, and their average accuracy is usually less than 10\%~\cite{kande2023llm}.

The main reasons for this dilemma include: 1. The existing framework didn't effectively suppress the inherent hallucination tendency or random generation characteristics of LLM~\cite{tambon2025bugs,lin2025explore}, which directly led to the great volatility of the accuracy and coverage of SVA; 2. The existing framework ignored the key role of the attention mechanism, which made LLM tend to focus on redundant or irrelevant information when processing, thus significantly reducing the accuracy; 3. LLM showed obvious limitations in processing formal languages and temporal logic. 4. The task is too complex, causing LLM to have errors superimposed; 5. LLM lacks a deep understanding of the syntax structure of the hardware description language (HDL), making it difficult to accurately infer the hierarchy and signal dependencies of the hardware design. To address these challenges, this paper proposes an innovative approach \TecName. Our main contributions are summarized as follows:

\begin{figure*}[ht]
\centering
\includegraphics[width=0.9\textwidth]{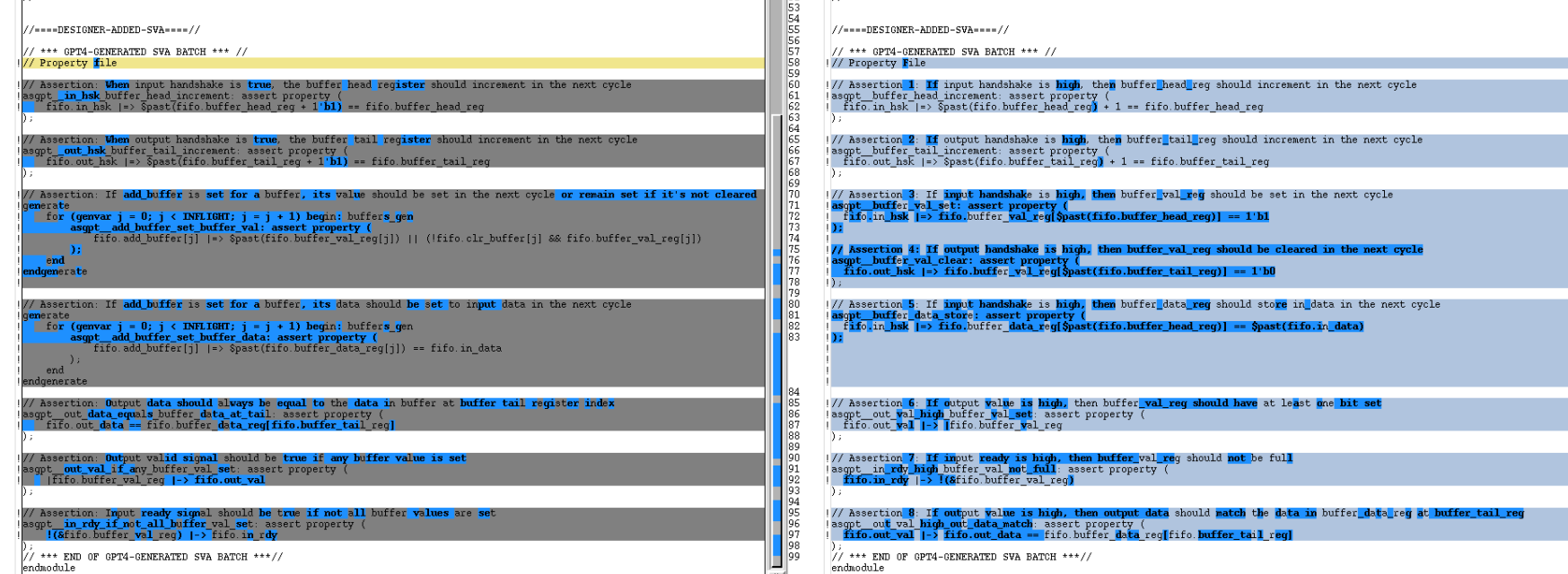}
\caption{There are many differences in the SVA generated by AutoSVA2 for the same design. This is the result of the LLM's attention mechanism, hallucination, and random answers. This makes the generated code much less trustworthy.}
\label{fig:autosva_diff}
\end{figure*}

\begin{itemize}
    \item We propose an SVA generation framework \TecName~based on fine-grained prompting techniques. Without requiring extensive training data or GPU/TPU clusters, high-quality SVAs that are syntactically and logically correct can be generated. 
    \item \TecName~is a general framework that can generate high-accuracy SVA when different LLM cores are applied. The experimental results show that \TecName~can effectively suppress the effects of hallucinations and random generation compared to other frameworks, and significantly improve the accuracy of the generated SVA.
    \item \TecName~can significantly reduce the required labor cost. We tested it on a bunch of benchmarks and the results show that \TecName~can reduce the workload of professional engineers.
    \item We further apply the \TecName~on one of the most efficient hardware vulnerability tools, SoFI~\cite{wang2021sofi}. The experiments demonstrate that \TecName~is highly scalable.
\end{itemize}

\section{Background} 
\label{sec:background}

\subsection{Hardware Security Oriented Assertion}
\label{sec:securityassertion}

The key step for hardware security verification is to proactively consider security threat modeling in the design lifecycle. In view of the diverse attacks commonly faced by modern hardware systems (including but not limited to non-invasive attack methods such as side-channel attacks\cite{ge2018survey,lou2021survey,he2022side,tehranipoor2023side,park2019leveraging} and fault injection attacks\cite{bhunia2018hardware,helfmeier2013breaking,tajik2015laser,zussa2013power,agoyan2010clocks}), a defensive design paradigm based on hardware description language (HDL) coding specifications needs to systematically implement security vulnerability mitigation strategies in the register transfer level (RTL) design phase. In view of this, the core feature of assertions used to verify security properties, which is different from traditional functional verification assertions, is that they focus on the formal representation of attack paths and vulnerability exploitability assessment. Specifically, by building a threat model-driven security property check database, potential security properties in hardware design can be identified, such as sensitive information leakage channels, unprotected ports, critical state machine transition vulnerabilities, etc. Most research is focused on the automatic generation technology of functional SVA, while the exploration of the field of automatic generation of security SVA is relatively scarce. Even in the existing SVA automatic generation technology, there are still a series of technical challenges that need to be solved: frequent syntax errors in the generation process lead to low verification efficiency, the inability to suppress the impact of LLM hallucinations on the results, and a large amount of engineer resources are required for result verification and adjustment.

\subsection{Automatic SVA Generation}
\label{sec:llm_for_codegen}

To simplify the process of assertion generation and reduce manual effort, automated assertion generation techniques have emerged and become increasingly popular~\cite{hertz2013mining, ahmed2021automap}. Generating SVA based on security properties can be essentially regarded as a complex natural language processing task, and LLMs have shown significant application potential in this field. Since the launch of ChatGPT, using LLM to assist designers in coding has developed into a widely recognized emerging method~\cite{thakur2024verigen}. The advantages of LLM in the field of SVA generation are mainly reflected in four key dimensions: First, through large-scale pre-training of massive text and code data, LLM has acquired rich programming language syntax and semantic knowledge, laying the foundation for accurately generating formal expressions that meet the requirements of specific fields; second, LLM has excellent context understanding ability and can accurately transform abstract natural language security requirements into formal expressions with strict rules; third, LLM's built-in reasoning mechanism enables it to effectively handle the complex temporal logic relationships and conditional judgments involved in SVA; finally, LLM's powerful pattern recognition and generalization capabilities enable it to extract key patterns from existing code examples and flexibly apply them to new verification scenarios, thereby significantly improving the automation and efficiency of security verification.

\subsection{Limitations of AI-Assisted SVA Generation}
\label{sec:limitations_llm_sva}

Despite the promising potential of LLMs in code generation, their application in the SVA generation process, especially in generating property assertions for hardware security, still faces some challenges and limitations. Firstly, the quality of SVA generated by existing LLM is low~\cite{kande2023llm}, with certain deficiencies at the logic and syntax levels, which leads the code to fail to pass relevant checks~\cite{pearce2022asleep}. This requires extensive manual review and correction work by engineers, affecting the efficiency of assisted programming. Secondly, while fine-tuning an LLM can enhance its performance in specific domains, this requires a large quantity of high-quality Verilog datasets and hardware resources as support~\cite{thakur2024verigen}. Unfortunately, due to the lack of public large-scale high-quality Verilog datasets, existing research often has to rely on limited-scale, template-generated datasets for training, and it is difficult for small and medium-sized enterprises with limited funds and technical reserves to meet hardware requirements. In addition, existing LLM-based methods and frameworks have significant deficiencies in code trustworthiness, and this problem is particularly serious in the field of hardware security verification. Specifically, for exactly the same design specifications and security requirements, the code generated by these methods in repeated experiments often shows worrying variability and unpredictability. This is the result of the combined effects of LLM hallucinations~\cite{tambon2025bugs}, unexpected/random answers~\cite{lin2025explore}, and LLM's attention mechanism~\cite{ben2024attend,tambon2025bugs,lin2025explore}. For natural language, the meaning can be understood by humans as long as it is expressed correctly, so there can be many changes in the details. But for code, a slight difference may cause a series of subsequent errors. As a key method for formal verification, the generation process of SVA must reduce the impact of LLM hallucinations and random answers and maintain a high degree of determinism, because any slight changes or logical differences may lead to inconsistent verification results, making the security vulnerability detection process unreliable. As shown in Figure~\ref{fig:autosva_diff}, for repeated experiments of the same design, there are many inconsistencies in the SVA generated by AutoSVA2. This makes the generated code much less trustworthy.

\begin{figure*}[htbp]
\centering
\includegraphics[width=0.9\textwidth]{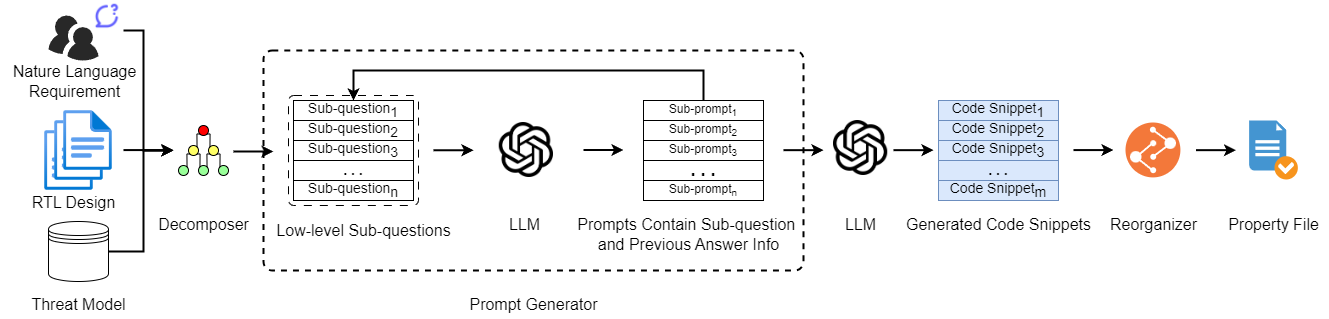}
\caption{\TecName~Powered LLM-based SVA Auto-Generation Flow.}
\label{fig:work_flow}
\end{figure*}

\section{Proposed Framework: \TecName} 
\label{sec:proposed_scheme}

To improve the performance of LLMs in specific domains, researchers have mainly explored two methodological paths: fine-tuning and prompt engineering. Model fine-tuning involves further adjusting and optimizing model parameters based on pre-trained models using large datasets in specific domains~\cite{lv2023full}. Although this method can theoretically significantly improve the performance in target domains, its implementation faces severe challenges: first, it requires the deployment of large-scale high-performance computing hardware resources; second, it places extremely high demands on the quality of training data; and finally, hyperparameter setting experience. The data must not only meet the model learning requirements in terms of scale and diversity but also must maintain a high degree of professionalism and accuracy. Any data quality issues and hyperparameter settings may lead to negative optimization of model performance. This high threshold means that for most companies with limited resources, it is almost impossible to meet both hardware and data collection requirements~\cite{xia2024understanding, lin2024data}. In contrast, prompt engineering, as an alternative method, guides the model to generate the expected output by carefully designing prompt statements, avoiding the complexity of directly modifying the internal structure or parameters of the model. This method not only has a low implementation threshold and high cost-effectiveness but also has significant flexibility~\cite{white2023prompt}. Researchers can quickly iterate and adjust the prompt strategy according to different application requirements so that the output of LLM can more accurately match specific scenarios and task requirements, and optimize model performance under resource-constrained conditions. Therefore, we developed \TecName~based on Prompt Engineering. \TecName~can effectively suppress LLM hallucinations and random answers, and generate high-quality SVA for hardware security. The subsequent chapters will describe each component of \TecName~in detail(Sec.\ref{sec:decomposer} - Sec.\ref{sec:reorganizer}).

\begin{algorithm}
\caption{\TecName}
\label{alg:decomposer_promptgenerator_reorganizer}
\footnotesize
\begin{algorithmic}[1]
\Procedure{SVA\_Gen\_Flow}{}\\
{\bf Input:} \textit{Requirement$\colon$str, RTL\_Design$\colon$list, Threat\_Model$\colon$list}\\
{\bf Output:} SVA
\State $Initialize()$
\State {\bf for} {$design$} {\bf in} {$RTL\_Design$}
\State \ \ {\bf for} {$threat$} {\bf in} {$Threat\_Model$}
\State \ \ $//\ decompose\ sub\_questions\ according\ to\ threat\ model$
\State \ \ \ \ $lSub\_Questions\ =\ Decomposer(design, threat)$
\State \ \ \ \ {\bf for} {$sub\_question$} {\bf in} {$lSub\_Questions$}
\State \ \ \ \ $//\ generate\ prompts\ for\ each\ question\ and\ get\ answer$
\State \ \ \ \ \ \ $strPrompt\ =\ PromptGen(sub\_question, strInfo, k)$
\State \ \ \ \ \ \ $strAnswer\ =\ AIAgent\_submit(strPrompt)$
\State \ \ \ \ \ \ $strInfo\ =\ extract\_info(strAnswer)$
\State \ \ \ \ \ \ $lAssetInfo\ +=\ extract\_asset(strInfo)$
\State \ \ \ \ {\bf end\ for}
\State \ \ \ \ {\bf for} {$asset$} {\bf in} {$lAssetInfo$}
\State \ \ \ \ $//\ generate\ sva\ code$
\State \ \ \ \ \ \ $strPrompt\ =\ PromptGenerator(SVAGenReq, asset)$
\State \ \ \ \ \ \ $lCodeSnippet\ +=\ AIAgent\_submit(strPrompt, asset)$
\State \ \ \ \ {\bf end\ for}
\State \ \ {\bf end\ for}
\State {\bf end\ for}
\State $//\ add\ module\ and\ ports\ info\ and\ reorganize$
\State $SVA\ =\ Regonizer(lCodeSnippet)$
\State \textit{{\bf return} SVA}
\EndProcedure
\end{algorithmic}
\end{algorithm}

\subsection{Overview}
\label{sec:Overview}

The workflow of \TecName~is illustrated in Figure~\ref{fig:work_flow} and Alg. \ref{alg:decomposer_promptgenerator_reorganizer}.
The framework begins by receiving three inputs: the RTL design to be verified, predefined threat models, and requirements in natural language(Line 2). The predefined threat models for hardware security mainly come from Trust-HUB~\cite{trusthub} and Common Weakness Enumeration (CWE)~\cite{cwe}. These inputs provide the necessary context and constraints for the subsequent decomposition and generation processes. One of the key components of \TecName~is the \textbf{Decomposer} module(Line 8). Due to the limited logical reasoning ability of LLMs~\cite{arkoudas2023gpt,lee2023mathematical}, the results of directly generating SVA based on circuits are often unsatisfactory, so we propose Decomposer. Its main task is to progressively decompose original security requirements into a series of finer-grained, more specific sub-questions. This decomposition process adheres to predefined threat models to ensure that the generated sub-questions encompass all relevant security aspects. The sub-questions are logically ordered from basic design information retrieval to the final generation of SVA code, forming a progressive problem-solving chain or reasoning chain. Each sub-question generated by the decomposer is then processed by the prompt generator (another key component of \TecName), which generates input prompts specific to that sub-question(Line 9 - Line 14). The purpose of the problem-solving chain is to guide LLM to gradually remove redundant information in complex designs and effectively focus LLM's attention on critical assets and threat models. Then, the highly refined information obtained is input into LLM to generate SVA snippets(Line 19). These code snippets are the input for the \textbf{Reorganizer} module. Once all sub-questions are processed, the Reorganizer module reassembles and integrates all generated code snippets according to predefined organizational rules, and through proper code organization and formatting, ultimately outputs a structurally complete SVA file(Line 24 - Line 25). The core idea of \TecName~is built on a key insight: complex goals can be broken down into smaller, more manageable sub-goals so that LLM can process them efficiently~\cite{wang2024survey,weng2023llm}.

\subsection{Decomposer}
\label{sec:decomposer}

The Decomposer plays a key role within the \TecName~framework, according to Alg. \ref{alg:decomposer_promptgenerator_reorganizer}, with its core function being to decompose highly abstract original requirements into a series of fine-grained sub-questions. This is also the part that requires engineers to participate. Directly using highly generalized natural language descriptions often leads to errors or inconsistencies in the answers generated by LLMs. This approach significantly reduces the complex reasoning required by LLMs, thus enabling the generation of more accurate, stable, and reliable answers.

\begin{figure}[h]
\centering
\includegraphics[width=0.95\columnwidth]{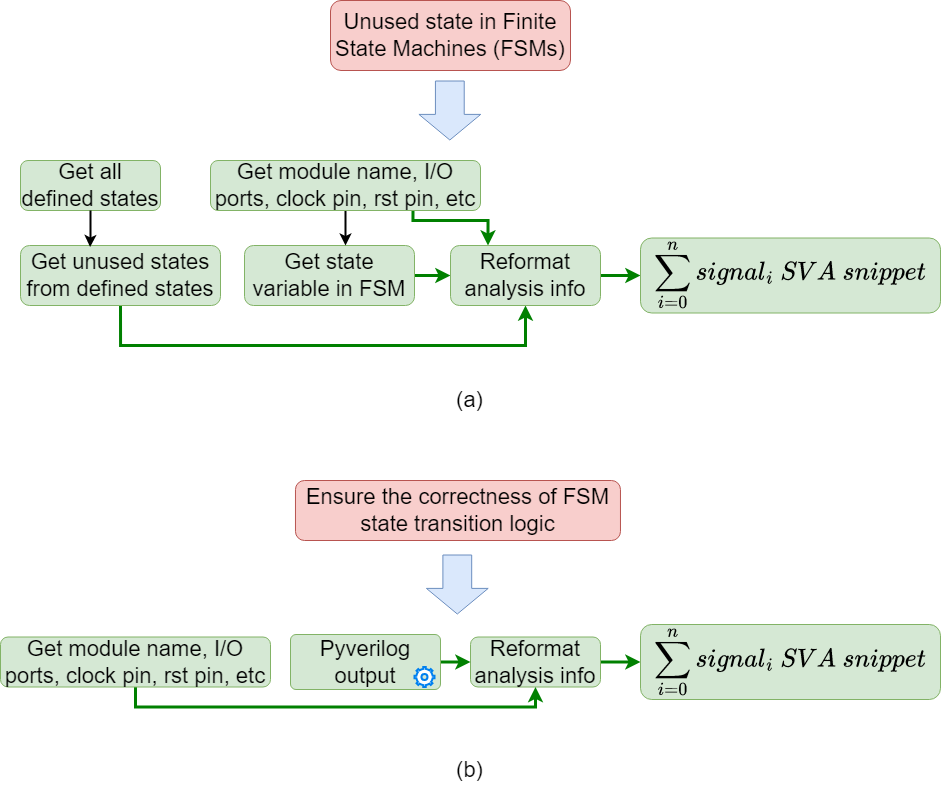}
\caption{Decomposer examples concerning (a) unused states in FSM, and (b) state transition logic in FSM. For (b), it uses part of the output from the third-party tool Pyverilog as input information. The black arrows ``\textbf{$\longrightarrow$}" represent the decomposed sub-questions, and the green arrows ``\textbf{$\textcolor[RGB]{53,126,22}{\longrightarrow}$}" represent the logical sequence.}
\label{fig:decomposer_example}
\end{figure}

Figure~\ref{fig:decomposer_example} shows how Decomposer decomposes the original requirements into fine-grained sub-questions for two different threat models. For different threat models, the decomposed sub-questions are also different. Figure~\ref{fig:decomposer_example}(a) shows the decomposition process to solve one of the common security vulnerabilities ``unused state (CWE-1245)" in FSM (finite state machine). We assume that all SVA snippets follow the following form:

\begin{equation}
\footnotesize
\label{eq_sva_format}
\begin{aligned}
&\textbf{property}\ property\_name\\
&\ \ \ \ \textbf{@(}edge\ module.senselist\textbf{)}\ \textbf{Left\_part}\ Operator\ \textbf{Right\_part;}\\
&\textbf{endproperty}\\
&def\_assert: \textbf{assert property}(property\_name);
\end{aligned}
\end{equation}

In order to make the generated SVA satisfy the format of Eq.~\ref{eq_sva_format}, LLM first needs to extract basic information from the circuit, including module name, I/O port, clock pin and reset pin. Then LLM needs to identify all defined states. Subsequently, based on the collected information, LLM needs to accurately identify the unused states and their related variables from FSM. Finally, all the valid information previously obtained is reformatted and the corresponding expression is generated for each unused state to meet the format in Eq.~\ref{eq_sva_format}. Repeat this step until all assets found by LLM are traversed. Figure~\ref{fig:decomposer_example}(b) shows an example of analyzing the state transition logic in FSM (RCD-011). Similar to the example in Figure~\ref{fig:decomposer_example}(a), this process also starts with a highly abstract requirement. In the FSM analysis phase, unlike Figure~\ref{fig:decomposer_example}(a), we embedded a third-party tool Pyverilog~\cite{takamaeda2015pyverilog} for analysis, and then passed its output information to LLM for processing. Finally, the corresponding SVA code snippet is generated for each state transition condition, and the process is looped until all relevant signals are covered. For any threat model, the idea of decomposing sub-questions is roughly the same. First, engineers need to know the expected format of SVA. Then, specific sub-questions are determined based on the expected format.

\subsection{Prompt Generator}
\label{sec:prompt_generator}

The next step in \TecName~involves the Prompt Generator creating prompt texts for each sub-question. These prompts ensure that the output from the LLM is both as expected and controlled. As described in Alg. \ref{alg:decomposer_promptgenerator_reorganizer}, the sequence of the sub-question prompted to LLM is crucial, as there are logical dependencies among them. For example, the sub-question, ``generating SVA code snippets" shown in Fig. \ref{fig:decomposer_example}(a), should be placed at the end of the solution chain because it relies on essential information collected from other sub-questions. Specifically, it is necessary to determine which module the target signal belongs to, clock pin, reset pin, and other information, so we put obtaining basic information before ``generating SVA code snippets". 

\begin{figure}[h]
\centering
\includegraphics[width=0.95\columnwidth]{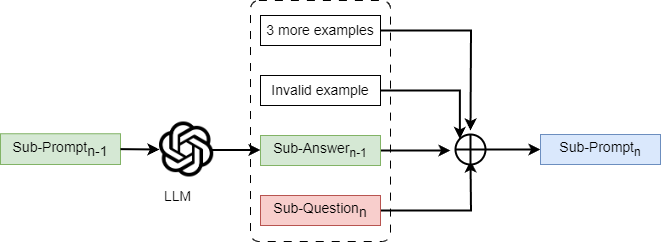}
\caption{The prompt generation for $sub-Question_n$ consists of three valid examples for $sub-Question_n$, one invalid example for $sub-Question_n$, $sub-Question_n$ itself, and valid information from the answer of $sub-Question_{n-1}$.}
\label{fig:prompt_generator}
\end{figure}

Figure~\ref{fig:prompt_generator} shows the structure of the generated prompt. For $sub-Question_n$ in the problem-solving chain, the Prompt Generator first considers the response of the LLM to $sub-Question_{n-1}$ to ensure the current prompt includes the necessary background information. According to Alg. \ref{alg:decomposer_promptgenerator_reorganizer}, we take the default value of k as 3 in our framework, and the Prompt Generator includes three valid examples for $sub-Question_n$, which demonstrates the expected input-output pattern, effectively guiding the LLM to generate responses that meet the requirements. Additionally, it also includes one invalid example to demonstrate how to handle exceptions, ensuring the LLM does not generate random inferences when encountering invalid inputs. After providing context and examples, the Prompt Generator clearly presents the actual $sub-Question_n$ at the end of the prompt, ensuring the LLM accurately understands the current task.

\begin{figure}[h]
\centering
\includegraphics[width=1\columnwidth]{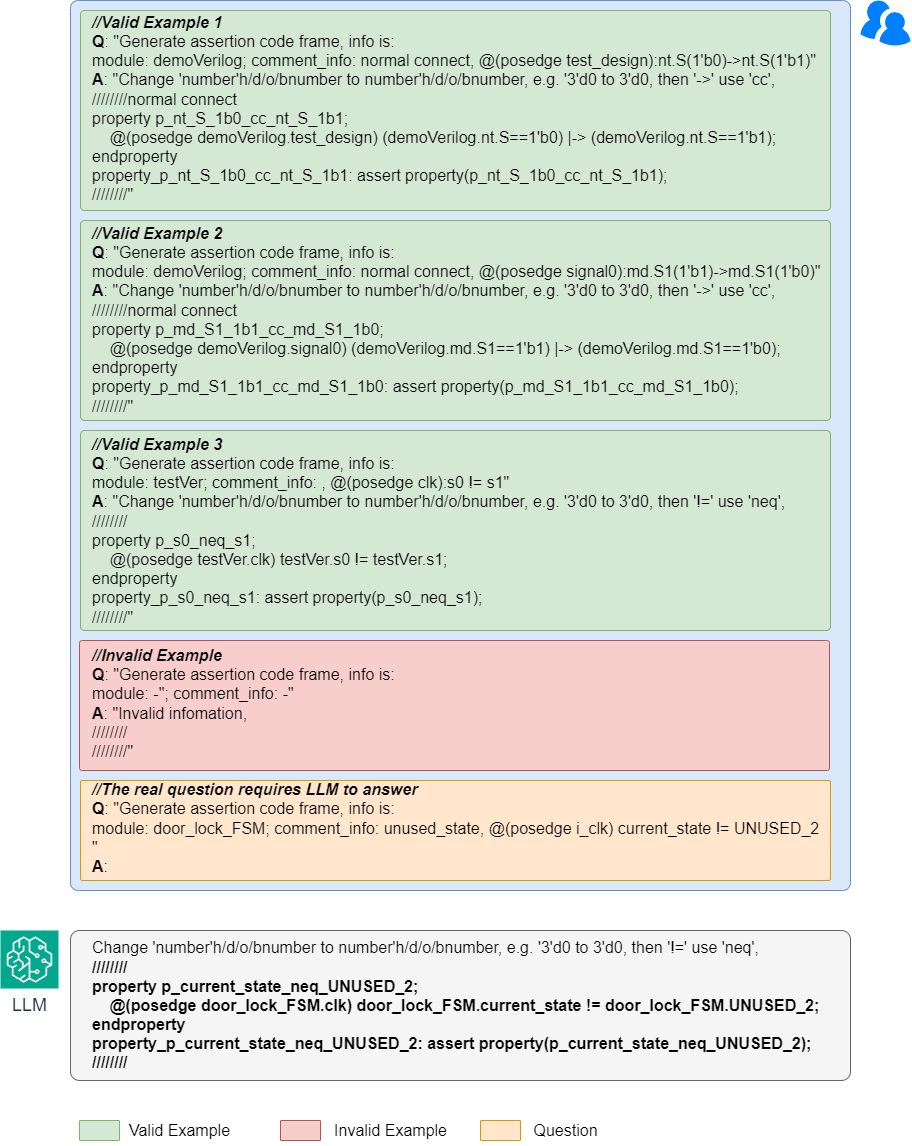}
\caption{The last sub-question of asset $current\_state$ to generate SVA for the unused state. The prompt contains 3 valid examples (green blocks) and 1 invalid example (red block), and the question (orange block) contains the key information obtained from the previous sub-questions. LLM only needs to complete the answer according to the example in the prompt.}
\label{fig:prompt_code_gen}
\end{figure}

Figure \ref{fig:prompt_code_gen} shows an example of the prompt used in the code generation stage. After obtaining all necessary design information and intermediate analysis results, the final step is to generate specific SVA code snippets in a structured and direct manner. The design of this prompt is consistent with the structure described in Figure~\ref{fig:prompt_generator}, which includes three typical valid input examples that demonstrate the correct input format and expected output for different scenarios. It also includes one invalid input example to clearly guide LLM in avoiding incorrect or random answers when faced with invalid problems or inputs. In the actual problem section, the prompt integrates key information obtained from the analysis of previous sub-questions, including the target module's name, a list of sensitive signals to monitor, specific signals of particular interest, and simplified Verilog statements, among other core content. Providing LLM with only the key information and context directly relevant to the current task significantly improves LLM’s reasoning accuracy. This is because LLM can focus its attention on more specific and explicit issues rather than trying to extract and understand all relevant details from a vast amount of design information. And then we can see that LLM outputs the expected answer.

\subsection{Reorganizer}
\label{sec:reorganizer}

In the previous steps, we obtained the problem-solving chain and generated prompt texts for all sub-questions. In addition, we have generated corresponding SVA code segments for each asset. The next task is to integrate all the generated code segments to create a directly usable SystemVerilog (SV) file. Overall, we have obtained all the elements needed to generate the SV file in the previous steps. This includes the wrapper of the file, which contains the module name, port definitions, and basic information such as clock and reset, binding functions, etc. We also obtained content needed for intermediate processes, such as assertion code segments corresponding to each target signal. Therefore, this step no longer relies on the reasoning capabilities of the LLM but is completed through an automation script running locally to generate the SystemVerilog file.

\section{Experimental Results} 
\label{sec:results}

We use JasperGold as the model checker tool and conduct experiments on nearly 500 designs from TRUST-HUB, Pyverilog, and~\cite{saha2024empowering}. All experiments are run on a server equipped with a 32-core 2.6 GHz Intel Xeon CPU and 64GB RAM.

\begin{figure}[h]
\centering
\includegraphics[width=1.0\columnwidth]{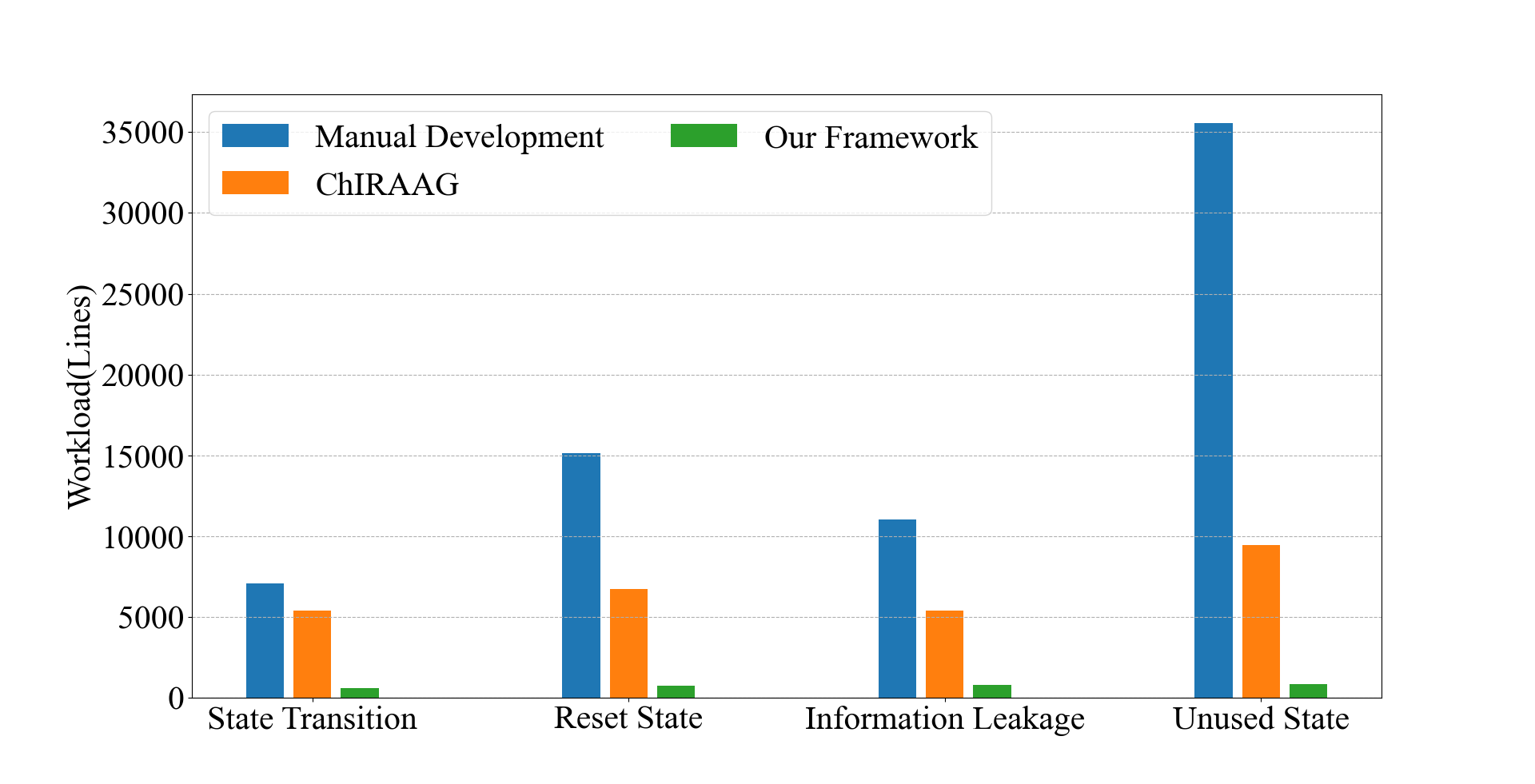}
\caption{Comparison of the amount of work required by engineers to write security property SVA for 450 circuit designs. Unlike other frameworks, \TecName~does not require re-design prompts for different designs.}
\label{fig:workload}
\end{figure}

\subsection{Engineer Workload}
\label{sec:experiment_engineer_workload}

From the perspective of engineering resource investment, existing automated SVA generation frameworks such as AutoSVA2~\cite{orenes2023rtl} or ChIRAAG~\cite{mali2024chiraag} have significant scalability limitations because they require engineers to redesign and customize prompt content for each different hardware design, whether these prompts are embedded in the form of code comments or expressed in structured JSON format. This feature makes these frameworks acceptable when dealing with a small number of verification tasks, but when faced with large-scale hardware design verification requirements, the workload of engineers will increase linearly, thus becoming a bottleneck for practical applications. Figure~\ref{fig:workload} clearly shows the comparison between \TecName~and existing methods in terms of engineer workload investment. In our analysis of 450 different circuit designs, the number of lines of code written by engineers is used as a quantitative metric of workload. The results show that the method adopted by \TecName~has a significant workload advantage. Specifically in this experimental environment, for all 450 circuits, the actual work of engineers is limited to the initial stage of writing prompt templates for each threat model, and these completely consistent prompt templates can be successfully applied to all designs to be verified. Engineers only need to write 500-700 lines of prompts for each threat model when utilizing \TecName. More importantly, as the number of circuits increases, \TecName's advantage in workload becomes more and more obvious, which has important practical value for large-scale hardware security verification projects.

\subsection{Accuracy and Universality}
\label{sec:experiment_universality_accuracy}

Universality is the key feature of \TecName, which enables the SVA generated by different LLMs to have high accuracy without the need for specialized training or fine-tuning. The core advantage lies in its model-agnostic nature, allowing researchers and engineers to use a variety of available LLMs to perform security SVA generation tasks without being limited by the availability of specific models or their special requirements. To comprehensively evaluate the practical performance of the \TecName~in generating high-quality SVA, we conducted experiments using benchmarks provided by Pyverilog~\cite{takamaeda2015pyverilog} and TrustHUB~\cite{trusthub}. The benchmarks and the corresponding threat models are shown in Table~\ref{tab:benchmarks}. These tests covered four major hardware design vulnerabilities that are often overlooked or difficult to detect in actual hardware designs. Specifically, they are state transition errors, unused states, information leakage, and incorrect initialization. For each benchmark design, we used five LLMs to generate the corresponding SVA code: GPT-4~\cite{openai2023gpt}, Gemini-Pro~\cite{team2023gemini}, Claude3\cite{Anthropic2024claude3}, Meta-AI\cite{metaai}, and Copilot~\cite{mscopilot}.

\begin{table}[]
\caption{Benchmarks and the corresponding threat models.}
\label{tab:benchmarks}
\resizebox{\linewidth}{!}{
\begin{tabular}{ccccc}
\hline
                        & State Transition Errors & Incorrect Initialization & Information Leakage & Unused State \\ \hline
blocking                & \checkmark                &             &                     & \checkmark            \\
case                    & \checkmark                &             &                     & \checkmark            \\
deepcase                & \checkmark                &             &                     &              \\
generate\_instance      & \checkmark                &             &                     &              \\
instance\_empty\_params & \checkmark                &             &                     &              \\
statemachine            & \checkmark                &             &                     &              \\
vectoradd               & \checkmark                &             &                     &              \\
door\_lock\_fsm         &                  & \checkmark           & \checkmark                   & \checkmark            \\
fifo4                   &                  & \checkmark           & \checkmark                   & \checkmark            \\
simple\_spi             &                  & \checkmark           & \checkmark                   & \checkmark            \\
fsm                     &                  & \checkmark           & \checkmark                   & \checkmark            \\
write\_once\_register   &                  & \checkmark           & \checkmark                   & \checkmark            \\ \hline
\end{tabular}
}
\end{table}

\begin{figure}[h]
\centering
\includegraphics[width=0.9\columnwidth]{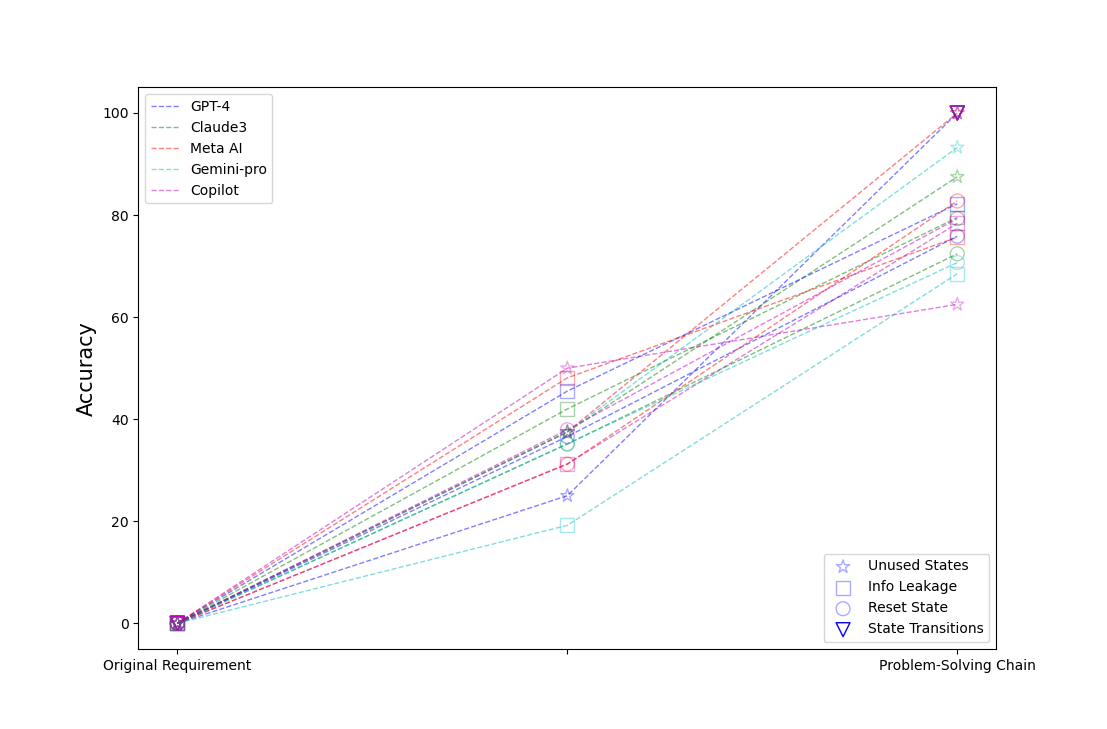}
\caption{Average correctness of generated SVA from original requirement to problem-solving chain.}
\label{fig:results_average_accuracy}
\end{figure}

\begin{table*}[]
\caption{Accuracy of generated SVA for different threat models.}
\label{tab:results}
\resizebox{\linewidth}{!}{
    \begin{tabular}{lccclccc}
    \hline
    Threat Model & \multicolumn{3}{c}{Unused States}                                                                                &                      & \multicolumn{3}{c}{Info Leakage}                                                                                 \\ \cline{2-4} \cline{6-8} 
    LLM          & \multicolumn{1}{l}{Avg. Sub-Questions} & \multicolumn{1}{l}{Functionality} & \multicolumn{1}{l}{Syntax} &                      & \multicolumn{1}{l}{Avg. Sub-Questions} & \multicolumn{1}{l}{Functionality} & \multicolumn{1}{l}{Syntax} \\ \cline{1-4} \cline{6-8} 
    GPT-4        & \multirow{5}{*}{9}                     & 100                               & 100                                 &                      & \multirow{5}{*}{10}                    & 82.2                              & 98.1                                \\
    Claude 3     &                                        & 87.5                              & 100                                 &                      &                                        & 79.5                              & 85.9                                \\
    Meta AI      &                                        & 100                               & 100                                 &                      &                                        & 75.7                              & 100                                 \\
    Gemini-Pro   &                                        & 93.3                              & 100                                 &                      &                                        & 79.3                              & 68.7                                \\
    Copilot      &                                        & 62.5                              & 100                                 &                      &                                        & 78.4                              & 82.8                                \\ \hline
    Threat Model & \multicolumn{3}{c}{Incorrect Initialization}                                                                                    &                      & \multicolumn{3}{c}{State Transition}                                                                             \\ \cline{2-4} \cline{6-8} 
    LLM          & \multicolumn{1}{l}{Avg. Sub-Questions} & \multicolumn{1}{l}{Functionality} & \multicolumn{1}{l}{Syntax} &                      & \multicolumn{1}{l}{Avg. Sub-Questions} & \multicolumn{1}{l}{Functionality} & \multicolumn{1}{l}{Syntax} \\ \cline{1-4} \cline{6-8} 
    GPT-4        & \multirow{5}{*}{7}                     & 75.9                              & 100                                 &                      & \multirow{5}{*}{12}                    & 100                               & 100                                 \\
    Claude 3     &                                        & 72.4                              & 95.2                                & \multicolumn{1}{c}{} &                                        & 100                               & 100                                 \\
    Meta AI      &                                        & 82.8                              & 91.7                                & \multicolumn{1}{c}{} &                                        & 100                               & 100                                 \\
    Gemini-Pro   &                                        & 70.7                              & 87.8                                & \multicolumn{1}{c}{} &                                        & 100                               & 100                                 \\
    Copilot      &                                        & 79.3                              & 100                                 & \multicolumn{1}{c}{} &                                        & 100                               & 100                                 \\ \hline
    \end{tabular}
}

\end{table*}

Figure~\ref{fig:results_average_accuracy} shows the average accuracy when using different LLMs to generate SVA. The figure shows that after decomposing the original requirements into problem-solving chains, the syntax and functionality accuracy show an upward trend compared to directly generating SVA. We use a scoring mechanism (out of 100 points) to evaluate the generated code, taking into account both functional/logical correctness and syntactic correctness. When using the original requirements directly, no model can generate completely correct, ready-to-use SVA code. This is entirely to be expected because generating SVA code directly is very challenging. LLM needs to take into account design details, syntax structure, timing, and other information at the same time. It is difficult to obtain satisfactory results for such a complex task. When we use \TecName, the generated SVA code achieves a high accuracy rate at both the logical and syntactic levels. This trend strongly proves that there is a significant positive correlation between the granularity of problem decomposition and the accuracy of SVA code generation. More importantly, as the decomposition granularity becomes finer, the ability of LLM to generate correct SVA code shows a steady upward trend. Between the original requirements and the problem-solving chain, although some models generate SVA that is completely correct at the logical and functional level, it still contains a large number of syntax errors. Therefore, it is necessary to add relevant sub-questions to correct syntax errors in the problem-solving chain.

Table~\ref{tab:results} shows the detailed accuracy of the generated SVA. In the table, $Avg.\ sub\-questions$ indicates the average number of sub-questions in the problem-solving chain. Since the number of assets in each circuit is different, the specific number of sub-questions will also vary. $Functionality$ indicates the logical/functional accuracy of the generated SVA, and $Syntax$ indicates the syntax accuracy of the generated SVA. It can be seen that the tested LLMs all have good accuracy performance. Especially for the threat model $State\ Transition$, since Pyverilog is embedded into the workflow to analyze the FSM, the accuracy of the generated SVA can reach 100\% on its highly refined results. We analyzed the reasons for the differences in SVA generated by different LLMs. Taking the performance of GPT-4 and Gemini-Pro as an example, GPT-4 is able to consistently generate SVAs with correct syntax and functionality for most of the threat models. In contrast, although the SVA generated by Gemini-Pro performs as well as GPT-4 in terms of logical functionality, it will have more syntax errors. For example, Gemini-Pro automatically adds extra backticks at the end of constant representations, mistakenly rewriting the standard Verilog constant representation $3'b001$ as $\textbf{3'b001{\color{red}'}}$. It is worth noting that all LLMs in these experiments used the same prompts, which indicates that the observed differences mainly stem from the intrinsic characteristics of the LLM itself and training biases. Gemini-Pro's performance seems to stem from its lack of recognition and processing of Verilog constants. It believes that quotation marks should appear in pairs, so it adds backticks after the constant $3'b001$ automatically. This finding explains why different LLMs performed similarly in terms of logical correctness, but the accuracy of the SVA code they generated was different. Similar problems also appeared in other models, and it included several different basic syntax errors such as clock/reset signal renaming and missing father modules. More importantly, such syntax errors can be effectively solved by introducing more sub-questions. For example, after generating the SVA Snippet, add the following sub-question and corresponding prompt examples: ``Check whether there are quotation marks at the end of all constants in the following SystemVerilog code snippet. If so, delete the quotation marks at the end."

\subsection{Hallucination Resistance and Scalability}
\label{sec:experiment_consistency_scalability}

\begin{figure}[h]
\centering
\includegraphics[width=1\columnwidth]{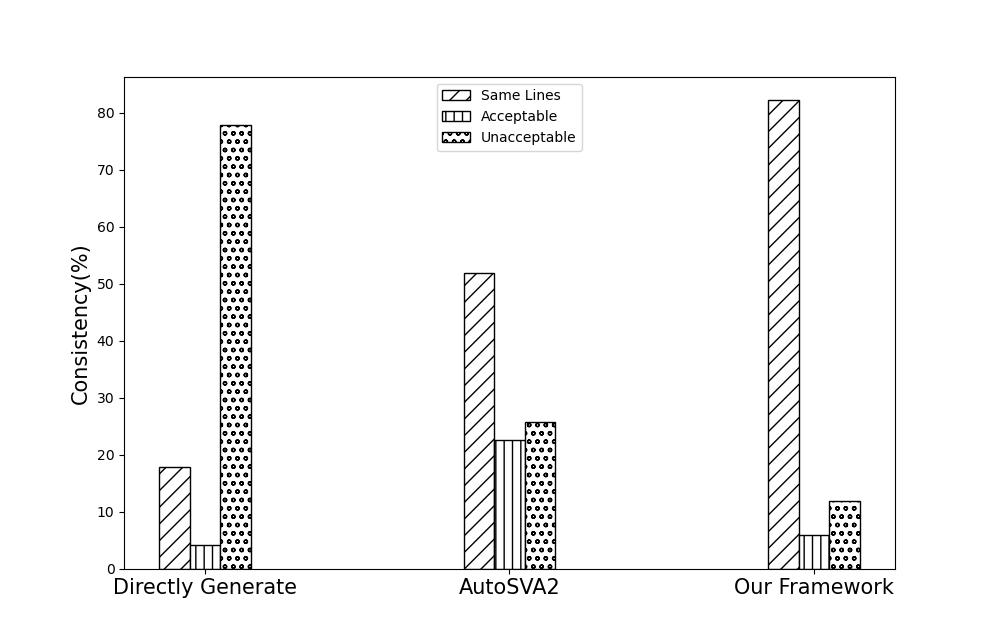}
\caption{Comparison of consistency of generated SVA. The smaller the impact of LLM hallucinations or unexpected/random answers, the higher the consistency of the generated SVA and the more trustworthy it is.}
\label{fig:results_consistency}
\end{figure}

In response to the impact of LLM hallucination and random answers proposed in Section \ref{sec:limitations_llm_sva}, we used the GPT-4 to test multiple benchmarks, and the results are shown in Figure~\ref{fig:results_consistency}. We test on 16 different circuits, and repeated the SVA generation experiment 5 times under the same conditions for each circuit. In the figure, $Same\ Lines$ means that the code generated in multiple experiments are same. $Acceptable$ means that although there are differences in the generated code, it does not affect reading and execution, such as partial differences in comments or different indentation and spaces used in the code. In this case, it often does not cause difficulties for engineers to understand the code, nor does it cause issues in EDA tools. $Unacceptable$ means that the generated code is difficult to understand or cannot be executed. It can be seen that the overall performance of \TecName~is better than AutoSVA2~\cite{orenes2023rtl}. \TecName~can reproduce more than 80\% of the code in multiple experiments, which effectively suppresses the influence of hallucinations and random answers, but there are still some cases where the code is unacceptable. These differences often occur when the circuit has too many IO ports. Even if \TecName~has removed redundant information, too many ports will still distract the effective attention of LLM. AutoSVA2 embeds hint comments in the design file, and then input both Verilog and comments together into the LLM for processing. This type of framework requires the LLM to complete two complex tasks at the same time: understanding the complex hardware design details and following various specific syntax rules, which is a very challenging task for the current LLM, and it is difficult to eliminate the influence of hallucinations and accidental/random answers.

We also test the performance of \TecName~when integrated with different tools. We focused on a framework: SoFI~\cite{wang2021sofi}. SoFI is a framework specifically designed to assess hardware vulnerabilities driven by security properties. One of the key steps in SoFI is that engineers are required to manually write security assertions based on security properties, a task that not only consumes considerable human resources but also requires engineers to have profound expertise and experience. In our experiments, we used \TecName~to automatically generate the corresponding SVA from SoFI's natural language descriptions, replacing the process of manually writing SVA. We selected two representative core security properties from SoFI for verification experiments, namely SP3.1 and SP3.2~\cite{wang2021sofi}.

\begin{itemize}
\item \textbf{SP 3.1:} At the 9th round of AES, any 1–3 Bytes of the first word in the round key cannot be faulty and the faulty bytes cannot propagate to the following words in the same round.
\item \textbf{SP 3.2:} At the 9th round of AES, 4 Bytes of any word in the round key cannot be faulty and the faulty bytes cannot propagate to the following words in the same round.
\end{itemize}
In this process, the sub-questions that LLM needs to answer include the arrangement of bytes, obtaining the specific number of bytes, and converting bytes to the corresponding bit range. LLM also needs to provide the byte arrangement, target bytes, and the final conversion results in a specific order. Finally, the correctness of the assertions generated for the intermediate state reached 91.67\%, and the correctness of the assertions generated by the round key reached 100\%. From the results, \TecName~can achieve satisfactory accuracy. More importantly, these results indicate that \TecName~has the potential to replace the process of manually writing SVAs in existing tools/frameworks, which not only greatly improves efficiency but also reduces the likelihood of human errors.

Currently, the SVA code generated by \TecName~has achieved high accuracy in functionality and syntax, but the generated SVA often follows similar patterns and structures. This is because we force the SVA generated by LLM to follow the format of Eq.~\ref{eq_sva_format}. The main differences between \TecName~and other LLM-based SVA generation frameworks are: 1) \TecName~no longer requires verification engineers to add explanatory comments about logic or requirements directly in the RTL design file, thereby avoiding the impact of the detail, style, and accuracy of the comments on the quality of the generated SVAs. 2) Verification engineers only need to participate once for a specific threat model and can apply it to different designs, effectively reducing the workload. 3) \TecName~effectively narrows the focus of LLM by dividing the sub-questions, thereby weakening the impact of hallucinations and random answers, making the generated SVA have a higher accuracy.

In practical hardware design and verification processes, the same functional requirement can often be described and implemented using various different SVA expressions. Although these expressions ultimately achieve equivalent effects, there may be differences in readability, maintainability, and resource usage. Therefore, empowering \TecName~to generate a variety of SVA expressions will help further enhance the applicability and practical value of the method.

\section{Conclusion} 
\label{sec:conclusion}

This paper proposes an innovative LLM-based general SVA generation framework, namely \TecName, for the field of hardware security. This method incorporates the concept of Prompt Engineering, guiding LLMs to generate high-quality SVA code that meets hardware verification requirements through finely-grained decomposed sub-questions. We considered various design vulnerabilities and combined this method with the existing SoFI framework to test \TecName~on popular LLMs. The experimental results show that \TecName~can achieve a high accuracy in syntax and functionality. In addition, \TecName~can effectively suppress the influence of LLM hallucination, thereby enhancing the consistency and reliability of verification.

\bibliographystyle{IEEEtran}
\bibliography{refs}

\end{document}